\documentclass[prb,showpacs,preprintnumbers,amsmath,amssymb,twocolumn,superscriptaddress,longbibliography]{revtex4-1}
\usepackage{graphicx}
\def\be{\begin{equation}}
\def\ee{\end{equation}}
\def\bea{\begin{eqnarray}}
\def\eea{\end{eqnarray}}

\begin{document}

\title{Ir 5$d$-band Derived Superconductivity in LaIr$_3$\\}
\author{A. Bhattacharyya}
\email{amitava.bhattacharyya@rkmvu.ac.in} 
\address{Department of Physics, Ramakrishna Mission Vivekananda Educational and Research Institute, Belur Math, Howrah 711202, West Bengal, India}
\affiliation{ISIS Facility, Rutherford Appleton Laboratory, Chilton, Didcot Oxon, OX11 0QX, United Kingdom} 
\author{D. T.  Adroja} 
\email{devashibhai.adroja@stfc.ac.uk}
\affiliation{ISIS Facility, Rutherford Appleton Laboratory, Chilton, Didcot Oxon, OX11 0QX, United Kingdom} 
\affiliation{Highly Correlated Matter Research Group, Physics Department, University of Johannesburg, PO Box 524, Auckland Park 2006, South Africa}
\author{P. K. Biswas} 
\affiliation{ISIS Facility, Rutherford Appleton Laboratory, Chilton, Didcot Oxon, OX11 0QX, United Kingdom} 
\author{Y. J. Sato}
\affiliation{Graduate School of Engineering, Tohoku University, Sendai 980-8577, Japan} 
\affiliation{Institute for Materials Research, Tohoku University, Oarai, Ibaraki 311-1313, Japan}
\author{M. R. Lees}
\affiliation{Department of Physics, University of Warwick, Coventry CV4 7AL, United Kingdom} 
\author{D. Aoki}
\affiliation{Institute for Materials Research, Tohoku University, Oarai, Ibaraki 311-1313, Japan} 
\author{A. D. Hillier} 
\affiliation{ISIS Facility, Rutherford Appleton Laboratory, Chilton, Didcot Oxon, OX11 0QX, United Kingdom}

\date{\today}

\begin{abstract}

We have studied the superconducting properties of LaIr$_3$ with a rhombohedral structure using magnetization, heat capacity, and muon-spin rotation/relaxation  ($\mu$SR) measurements. The zero-field cooled and field cooled susceptibility measurements exhibit a superconducting transition below $T_{\mathrm{C}}$ = 2.5 K. Magnetization measurements indicate bulk type-II superconductivity with upper critical field $\mu_0H_{\mathrm{c2}}(0)$ = 3.84 T. Two successive transitions are observed in heat capacity data, one at $T_{\mathrm{C}}$ = 2.5 K and a second at 1.2 K below  $T_{\mathrm{C}}$ whose origin remain unclear. The heat capacity jump reveals  $\Delta C$/$\gamma T_{\mathrm{C}} \sim$ 1.0 which is lower than 1.43 expected for BCS weak coupling limit. Transverse field-$\mu$SR measurements reveal a fully gapped $s-$wave superconductivity with 2$\Delta(0)/k_{\mathrm{B}}T_{\mathrm{C}}$ = 3.31, which is small compared to BCS value 3.56, suggesting weak coupling superconductivity. Moreover the study of the temperature dependence of the magnetic penetration depth estimated using the transverse field-$\mu$SR measurements gives a zero temperature value of the magnetic penetration depth $\lambda_{\mathrm{L}}(0)$ = 386(3) nm, superconducting carrier density $n_{\mathrm{s}}$ = 2.9(1) $\times$10$^{27}$ carriers $m^{-3}$ and the  carriers' effective-mass enhancement $m^{*}$ = 1.53(1) $m_{\mathrm{e}}$. Our zero-field-$\mu$SR measurements do not reveal the spontaneous appearance of an internal magnetic field below the transition temperature, which indicates that time-reversal symmetry is preserved in the superconducting state of LaIr$_3$.

\end{abstract}

\pacs{71.20.Be, 75.10.Lp, 76.75.+i}

\maketitle

\section{Introduction}

\noindent The quest for unconvetional superconductivity in materials based on 5$d$ transition metal elements is one of the most fascinating and significant topics in condensed matter physics~\cite{B. J. Kim,B. J. Kim1,S. J. Moon, Hirai2010}. Due to the presence of strong spin-orbit (SO) coupling effects, the 5$d$ transition metal compounds have been comprehensively investigated to find a correlation between strong SO coupling and unconventional superconductivity~\cite{J. H. Kim,G. Jackeli}. The close correspondence of comparatively weak electronic correlations (0.5-3 eV), strong crystal field effects (1-5 eV) and strong relativistic SO coupling effects (0.1-1 eV) offers an remarkably promising opportunity for the study of the physics resulting from competing spin, orbital, charge, and lattice degrees of freedom~\cite{Y. Singh,S. K. Choi}. For example, pressure induced superconductivity in 1T-TaS$_2$~\cite{B. Sipos}, in noncentrosymmetric CePt$_3$Si~\cite{E. Bauer}, and in the geometrically frustrated pyrochlore oxides Cd$_2$Re$_2$O$_7$~\cite{M. Hanawa} and KOs$_2$O$_6$~\cite{S. Yonezawa}. In iriduim-containing compounds, superconductivity has been reported in materials such as IrSe$_2$, Cu$_{1-x}$Zn$_x$Ir$_2$S$_4$, CeIrSi$_3$, ScIrP, LaIrP and LaIrAs, etc. and in the ternary ThCr$_2$Si$_2$-type compounds BaIr$_2$P$_2$ and SrIr$_2$As$_2$~\cite{J. G. Guo,H. Suzuki,I. Sugitani,Y. Okamoto,Y. P. Qi,N. Berry}. In the case of rare-earth Ir based superconductors most of their electronic characteristics are due to the rare earth components rather than Ir. Nevertheless, there are a few cases, such as CaIr$_2$~\cite{H. M. Tutuncu}, IrGe~\cite{D. Hirai} and Mg$_{10}$Ir$_{19}$B$_{16}$~\cite{T. Klimczuk}, where the superconductivity arises from Ir 5$d$ states at the Fermi surface. Until the discovery of LaIr$_3$, no simple example of a La-Ir superconductor has been found whose properties are controlled by the Ir 5$d$ bands at the Fermi level and with strong spin-orbit-coupling.

\begin{figure*}[t]
\centering
    \includegraphics[height=0.3\linewidth,width=\linewidth]{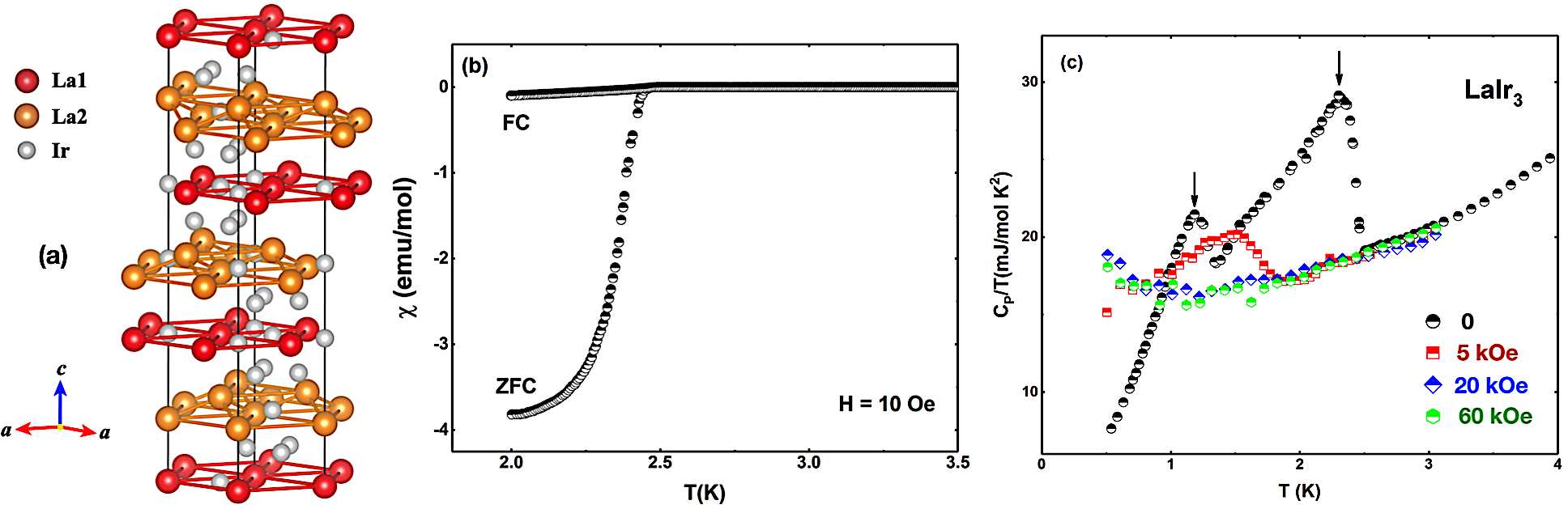}
\caption{(Color online) (a) 3D Crystal structure of rhombohedral LaIr$_3$. (b) Temperature dependence of the magnetic susceptibility of LaIr$_3$ in magnetic fields of 10 Oe in the zero-field cooled (ZFC) and field cooled (FC) states.~(c) $C_{\mathrm{P}}/T$ vs $T$ curves in different applied fields.}
\label{mtrtcp}
\end{figure*}

\noindent Recently Haldolaarachchige {\it et al.}~\cite{N. Haldolaarachchige} reported that LaIr$_3$ shows superconductivity with $T_{\mathrm{C}}$ = 2.5 K, where the bands near the Fermi surface are governed by the Ir 5$d$ states that are heavily influenced by spin-orbit coupling and there is no strong contribution from the La-orbitals near $E_{\mathrm{F}}$. LaIr$_3$ is therefore one of the few superconductors where 5$d$ electrons play the principal role in the appearance of superconductivity~\cite{N. Haldolaarachchige}. Furthermore, three-dimensional metallic character is seen from the band structure calculations; many bands with large dispersion cross $E_{\mathrm{F}}$~\cite{N. Haldolaarachchige}.  LaIr$_3$ crystallizes in the PuNi$_3$-type rhombohedral structure with the space group $R\bar{3}m$ (166, $D^5_{\mathrm{3d}}$ )~\cite{N. Haldolaarachchige}. There are two crystallographically inequivalent La sites and three Ir sites~\cite{N. Haldolaarachchige}. To examine the role of lanthanide elements and the spin-orbit coupling effects arising from the Ir 5$d$ bands in determining the superconducting properties, the study of other rhombohedral structure type $R$Ir$_3$ materials with $R$-lanthanides (with 4$f$)~\cite{J. G. Huber,Y. J. Sato} will be important. In particular, materials with partially filled 4$f$ orbitals, such as Ce, Pr, etc., may exhibit magnetism effects and possibly superconductivity. Sato {\it et al.}~\cite{Y. J. Sato} reported that CeIr$_3$ is a bulk type-II superconductor with a $T_{\mathrm{C}}$ = 3.4 K, which is the second highest $T_{\mathrm{C}}$ among the Ce-based intermetallic compounds. The low value of the electronic heat capacity coefficient, 23 mJ K$^{-2}$ mole$^{-1}$, shows CeIr$_3$ is a weakly correlated electron system~\cite{Y. J. Sato}.  

\noindent In this paper, we have investigated the superconducting properties of LaIr$_3$ by means of dc magnetization, heat capacity ($C_{\mathrm{P}}$), tranverse field muon spin rotation (TF$-\mu$SR), and zero-field muon spin relaxation (ZF$-\mu$SR) measurements. An analysis of the temperature dependence of the magnetic penetration depth measured using the TF$-\mu$SR measurements suggests an isotropic $s-$wave character for the superconducting gap. The  ZF$-\mu$SR result does not show spontaneous appearance of an internal magnetic field in the  superconducting state, suggesting time-reversal symmetry is preserved for LaIr$_3$.

\section{Experimental Details}

\noindent Polycrystalline samples of  LaIr$_3$ were prepared by arc melting of the constituent elements as reported by Haldolaarachchige {\it et al.}~\cite{N. Haldolaarachchige}. Powder X-ray diffraction measurement was carried out using a Panalytical X-Pert Pro diffractometer. Magnetic susceptibility measurements were made using a Magnetic Property Measurement System (MPMS) superconducting quantum interference device (SQUID) magnetometer (Quantum Design). Heat capacity measurements were performed by the relaxation method in a Quantum Design Physical Property Measurement System (PPMS). Temperatures down to 0.35 K were attained by a $^3$He insert in the PPMS.  We used the $\mu$SR technique to examine the superconducting ground state of LaIr$_3$. ZF$-\mu$SR measurements were performed on the MUSR spectrometer with the detectors in the longitudinal configuration at the ISIS Muon Facility located at the Rutherford Appleton Laboratory, United Kingdom. For the TF$-\mu$SR measurements the spectrometer were rotated by 90$^{\circ}$. The powdered sample of LaIr$_3$ was mounted on a high purity (99.995\%) silver plate using diluted GE varnish which was cooled down to 0.1 K using a dilution refrigerator. Using an active compensation system the stray magnetic fields at the sample position were canceled to a level of 1 mG. Spin-polarized muon pulses were implanted into the sample and the positrons from the resulting decay were collected in the detectors positions either forward or backward of the initial muon spin direction. For ZF$-\mu$SR measurements the asymmetry of the muon decay is calculated by, $G_{\mathrm{z}}(t) = [ {N_{\mathrm{F}}(t) -\alpha N_{\mathrm{B}}(t)}]/[{N_{\mathrm{F}}(t)+\alpha N_{\mathrm{B}}(t)}]$, where $N_{\mathrm{B}}(t)$ and $N_{\mathrm{F}}(t)$ are the number of counts at the detectors in the forward and backward positions and $\alpha$ is a constant determined from calibration measurements made in the normal state with a small (20 Oe) applied transverse magnetic field. The data were analyzed using the software WiMDA~\cite{FPW}.  

\begin{figure*}[t]
\centering
    \includegraphics[height=0.3\linewidth,width=\linewidth]{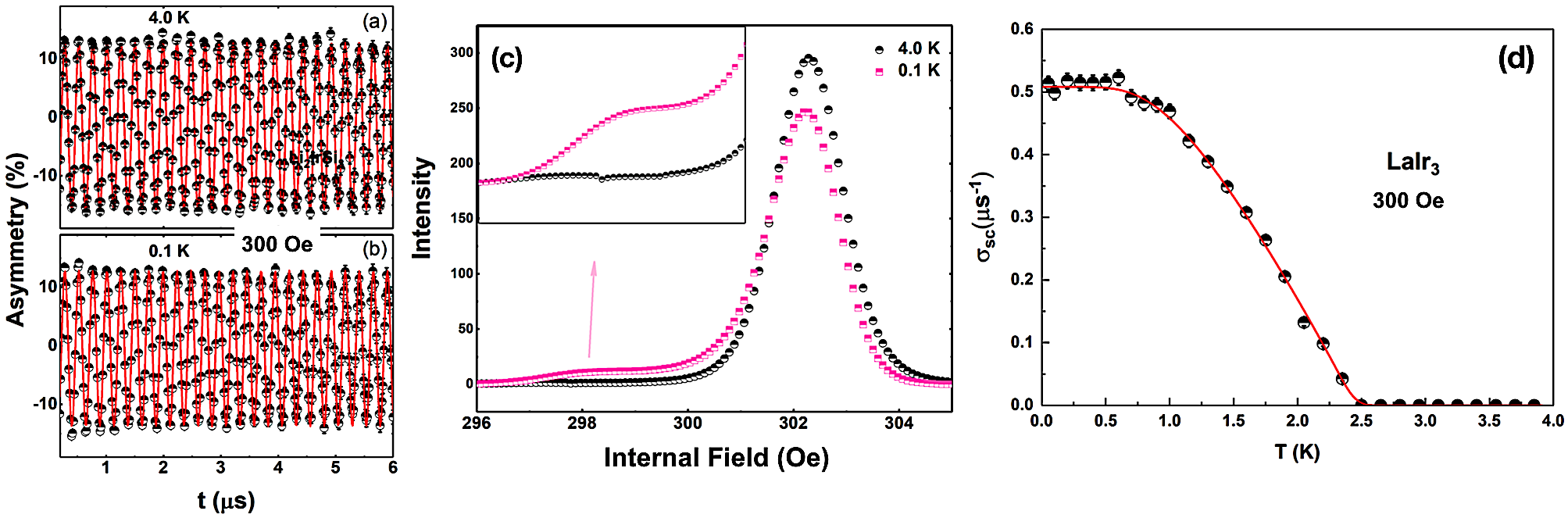}
\caption{(Color online) TF-$\mu$SR spin precession signals of LaIr$_3$ taken at applied magnetic field of $H$ = 300 Oe. Asymmetry vs time plot in (a) the normal state at 4.0 K and in (b) the superconducting state at 0.1 K. Solid lines represent fits to the data using Eq. (1). (c) Maximum entropy spectra (above and below $T_{\mathrm{C}}$). (d) Temperature dependence of the muon Gaussian relaxation rate $\sigma_{\mathrm{sc}}(T)$. The line is a fit to the data using an isotropic model (Eq. (2)).}
\label{gap}
\end{figure*}

\section{Results and Discussion}
\subsection{Crystal structures and physical properties}

\noindent Powder x-ray diffraction investigation confirmed that the sample is single phase with a rhombohedral crystal symmetry with the space group 166 ($R\bar{3}m$)~\cite{N. Haldolaarachchige}. Fig.~\ref{mtrtcp}(a) shows the crystal structure of LaIr$_3$. Red symbols represent La1 atoms, yellow symbols represent La2 atoms and small grey symbols represent Ir atoms.  A rhombohedral unit cell contains two distinct La atoms (La1 and La2) and three distinct Ir atoms (Ir1, Ir2, and Ir3).  The bulk nature of superconductivity in LaIr$_3$ was confirmed by the magnetic susceptibility $\chi(T)$, as shown in Fig.~\ref{mtrtcp}(b). The low field $\chi(T)$, measurements displays a strong diamagnetic signal due to a superconducting transition temperature  $T_{\mathrm{C}}$ = 2.5 K. The magnetization $M(H)$ curve (not shown here) at low temperature suggest type-II superconductivity. 

\par

\noindent Fig.~\ref{mtrtcp}(c) shows the $C_{\mathrm{P}}(T)/T$ at different applied magnetic fields. At 2.5 K, a clear anomaly is seen showing the superconducting transition which matches well with $\chi(T)$ data. We have observed a second transition in the heat capacity at 1.2 K, which is reminiscent of the two peaks in the heat capacity of UPt$_3$~\cite{UPt3}. Nevertheless, our TF-$\mu$SR studies do not show any clear changes at the lower transition and further investigations are needed in order to identify the origin of the second transition. As we do not see any change in ZF-$\mu$SR data at 0.09 K compare with 4 K, this also supports that second transition is not due to a magnetic impurity. In an applied magnetic field of 60 kOe, the heat capacity jumps fully suppressed. Since the normal-state specific heat was found to be invariant under external magnetic fields, the normal-state electronic specific heat coefficient $\gamma$ and the lattice specific heat coefficient $\beta$ were deduced from the data in a field of 60 kOe by a least-square fit of the $C_{\mathrm{P}}(T)/T$ data to $C_{\mathrm{P}}(T)/T = \gamma +\beta T^2$, which gives a Sommerfeld constant $\gamma$ = 15.32(3) mJ/(mol-K$^2$), $\beta$ = 0.56(1) mJ/(mol-K$^4$), and from this value of  $\beta$ (= ${nN_A}\frac{12}{5}\pi^4R\Theta_D^{-3}$, where $R$ = 8.314 J/mol-K, $n$ is the number of atoms per formula unit, and $N_A$ is Avogadro's number), we have estimated the Debye temperature $\Theta_D$ = 430(4) K. The value of $\gamma$ obtained is comparable to the cubic Laves phase superconductor CaIr$_2$, and some other Ir-based heavy element superconductors, which further suggests the common effect of Ir-sub-lattice on many Ir based superconducting materials. In the case of LaIr$_3$ the low value of $\gamma$ may be due to the absence of $f-$ orbitals near the Fermi level. Using the heat capacity jump we have calculated  the dimensionless parameter $\Delta C_{\mathrm{P}}$/$\gamma T_{\mathrm{C}} \sim$1 which is lower than 1.43 expected for BCS weak coupling limit. 

\begin{figure*}[t]
\centering
    \includegraphics[height=0.3\linewidth,width=\linewidth]{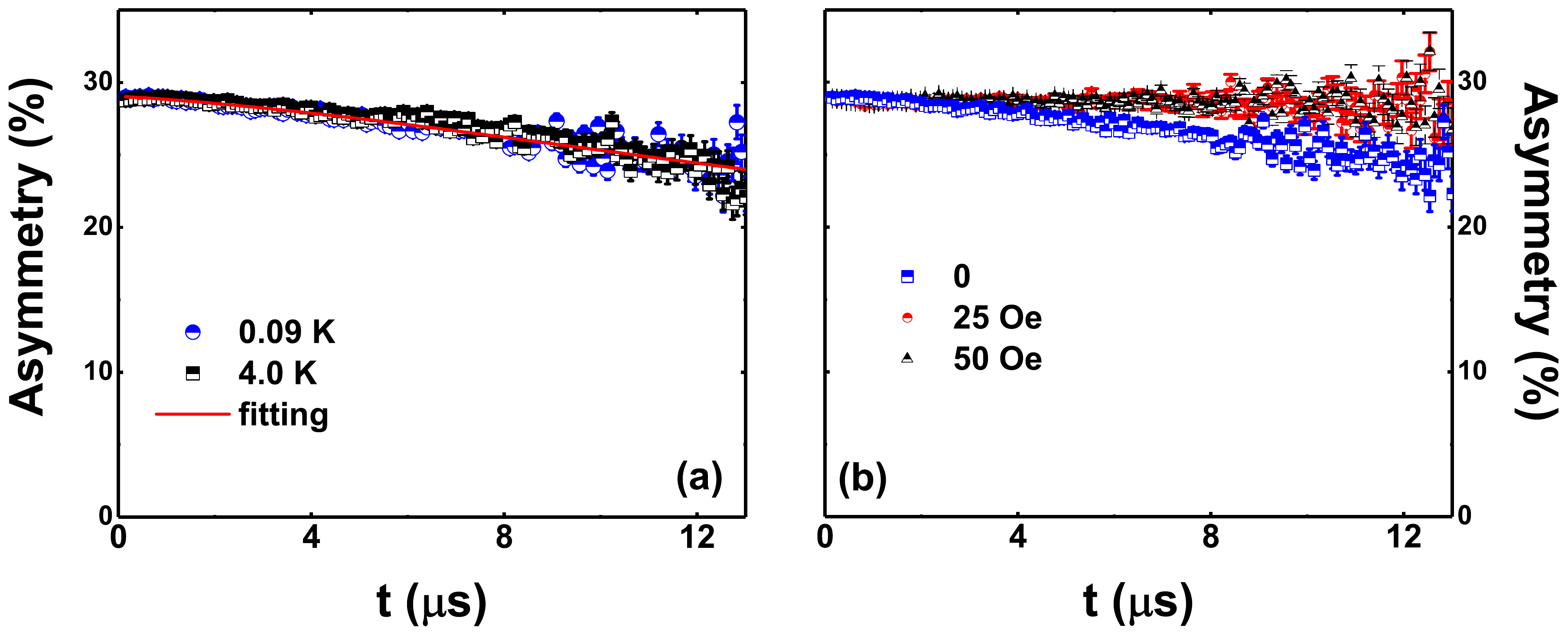}
\caption{(Color online) (a) Zero-field $\mu$SR time spectra for LaIr$_3$ collected at 0.09 K (squares) and 4.0 K (circles) are shown together with lines that are least squares fits to the data using Eqs. (4-5). (b)  Longitudinal field $\mu$SR time spectrum taken at 0.09 K in the presence of various applied magnetic fields.}
\label{ZFasymmetry}
\end{figure*}

\subsection{Superconducting gap structures}

\noindent Fig.~\ref{gap}(a-b) shows the  TF-$\mu$SR spectra above and below $T_{\mathrm{C}}$ and Fig.~\ref{gap}(b) shows the corresponding maximum entropy plots. It is clear that above $T_{\mathrm{C}}$ the $\mu$SR spectra show a very small relaxation mainly from the quasi-static nuclear moments, and the internal field distribution is very sharp and centered near the applied field. However at 0.1 K the $\mu$SR spectra show moderate damping and the internal field distribution has two components, one sharp near the applied field and one  broad which is shifted lower than applied field. In order to shed light on the pairing mechanism and the superconducting gap structure of LaIr$_3$ compound, we have examined the temperature dependence of the TF-$\mu$SR data.  Below $T_{\mathrm{C}}$ the asymmetry decay with time due to inhomogeneous field distribution of the flux-line lattice. TF-$\mu$SR asymmetry spectra at all temperatures below and above $T_{\mathrm{C}}$ could be best fitted by using an oscillatory decaying muon spin depolarization function~\cite{Bhattacharyya1,Bhattacharyya2,Adroja,Bhattacharyyarev},

\begin{equation}
G_{z1}(t) = A_{1}\cos(\omega_{1}t+\phi)\exp\bigg(\frac{-\sigma^{2}t^{2}}{2}\bigg)+A_{2}\cos(\omega_{2}t+\phi)
\end{equation}

\noindent where $A_{1}$ and $A_{2}$ are the transverse field asymmetries results from the sample and background from the Ag sample holder respectively, and $\omega_{1}$ and $\omega_{2}$ are the frequencies of the muon precession in the sample and background respectively, $\phi$ is the initial phase and $\sigma$ is the total Gaussian muon depolarization rate. The superconducting contribution to the muon depolarization rate $\sigma_{\mathrm{sc}}$ is estimated employing [$\sigma_{\mathrm{sc}} = \sqrt{\sigma^{2}-\sigma_{\mathrm{n}}^2}$], where $\sigma_{\mathrm{n}}$ is the nuclear contribution which is assumed to be fixed over the whole temperature range and was obtained by quadratically subtracting the background nuclear part obtained from the value estimated above $T_{\mathrm{C}}$. In the fitting of TF$-\mu$SR data given in Fig.~\ref{gap}(c), we kept $A_{2}$ fixed at its low-temperature value and this value was estimated first by fitting the lowest temperature data. $A_{1}$ was allowed to vary and its value is nearly temperature independent. The value of the phase was first estimated from the lowest temperature and kept fixed for all other temperatures. We can model the temperature dependence of the penetration depth/superfluid density using the following equation \cite{ThCoC2,Bhattacharyya1,Adroja} 

\begin{eqnarray}
\frac{\sigma_{\mathrm{sc}}(T)}{\sigma_{\mathrm{sc}}(0)} &=& \frac{\lambda^{-2}(T,\Delta_{0,i})}{\lambda^{-2}(0,\Delta_{0,i})}\\
 &=& 1 + \frac{1}{\pi}\int_{0}^{2\pi}\int_{\Delta(T)}^{\infty}(\frac{\delta f}{\delta E}) \times \frac{EdEd\phi}{\sqrt{E^{2}-\Delta(T,\Delta_{i}})^2} \nonumber
\end{eqnarray}

\noindent where $f= [1+\exp(-E/k_{B}T)]^{-1}$ is the Fermi function, $\phi$ is the angle along the Fermi surface, and $\Delta_{i}(T,0) = \Delta_{0,i}\delta(T/T_{c})g(\phi)$. The temperature variation of the superconducting gap is approximated by the relation $\delta(T/T_{\mathrm{C}}) = \tanh[{1.82(T_{\mathrm{C}}/T-1)]^{0.51}}$ where g($\phi$) refers to the angular dependence of the superconducting gap function and $\phi$ is the polar angle for the anisotropy. g($\phi$) is substituted by (a) 1 for an $s$-wave gap, (b) $\vert\cos(2\phi)\vert$ for a $d$-wave gap with line nodes~\cite{Annett,Pang}. The data can be well modeled using a single isotropic $s-$wave gap of 0.35(1) meV. This gives a gap ratio of 2$\Delta(0)$/k$_{B}$$T_{\mathrm{C}}$ = 3.31(1), which lower than the value as expected from BCS theory 3.56. This suggest weak coupling superconductivity in the case of LaIr$_3$ which is in agreement with heat capacity data. TF$-\mu$SR measurements of Mg$_{10}$Ir$_{19}$B$_{16}$ reveal spin-singlet $s-$wave pairing, even though the lack of inversion symmetry and the large SO coupling of Mg$_{10}$Ir$_{19}$B$_{16}$ should produce a mixed pairing state, likely with a large spin-triplet contribution~\cite{A. A. Aczel}. Furthermore, TF$-\mu$SR measurements on La$_7$Ir$_3$ suggest that the superconducting gap is isotropic and that the pairing symmetry of the superconducting electrons is predominantly $s-$wave with an enhanced binding strength~\cite{J.A.T. Barker}.

\noindent The observed large value of the muon spin depolarization rate below the superconducting transition temperature is related to the superfluid density or penetration depth. For a triangular~\cite{Sonier,Chia,Amato} lattice  $\frac{\sigma_{\mathrm{sc}}^2}{\gamma_{\mu}^2}=\frac{0.00371 \times \phi_{0}^{2}}{\lambda^4}$, where $\phi_{0}$ is the flux quantum number (2.07 $\times$10$^{-15}$Tm$^{2}$) and $\gamma_{\mu}$ is the muon gyromagnetic ratio $\gamma_{\mu}/2\pi$ = 135.5 MHz T$^{-1}$. As with other phenomenological parameters characterizing the superconducting state, the superfluid density can also be related to quantities at the atomic level. Using London's theory \cite{Sonier} $\lambda_{\mathrm{L}}^2 = \frac{m^{*}c^{2}}{4\pi n_{\mathrm{s}}e^{2}}$, where $m^{*} = (1+\lambda_{\mathrm{e-ph}})m_{\mathrm{e}}$ is the effective mass and $n_{\mathrm{s}}$ is the density of superconducting carriers. Within this simple picture, $\lambda_{L}$ is independent of magnetic field. $\lambda_{\mathrm{e-ph}}$ is the electron-photon coupling constant that can be estimated from the Debye temperature ($\Theta_{D}$) and $T_{\mathrm{C}}$ using McMillans relation \cite{McMillan}

\begin{equation}
\lambda_{\mathrm{e-ph}} = \frac{1.04+\mu^{*}\ln(\Theta_{D}/1.45T_{\mathrm{C}})}{(1-0.62\mu^{*})\ln(\Theta_{D}/1.45T_{\mathrm{C}})-1.04}
\end{equation}

\noindent Here $\mu^{*}$ is the repulsive screened Coulomb parameter with a typical value of $\mu^{*}$ = 0.15, give $\lambda_{\mathrm{e-ph}}$ = 0.53.  As LaIr$_3$ is a type II superconductor,  the magnetic penetration depth $\lambda$, superconducting carrier density $n_{\mathrm{s}}$, the effective-mass enhancement $m^{*}$ have been estimated to be $\lambda_{\mathrm{L}}(0)$ = 386(3) nm, $n_{\mathrm{s}}$ = 2.9(1) $\times$10$^{27}$ carriers $m^{-3}$, and $m^{*}$ = 1.53 $m_{\mathrm{e}}$ respectively, for LaIr$_3$.

\subsection{Zero-field muon spin relaxation}

\noindent Fig.~\ref{ZFasymmetry}(a) shows the time evolution of the zero-field muon spin relaxation asymmetry spectra in LaIr$_3$ at temperatures above and below $T_{\mathrm{C}}$. Below $T_{\mathrm{C}}$, we do not observe any change in the muon spin relaxation rate with decreasing temperature down to the lowest temperature (0.09 K) of measurements, which indicates the absence of any spontaneous internal field in the superconducting phase. The ZF$-\mu$SR spectra for LaIr$_3$ can be well described by the damped Gaussian Kubo-Toyabe (K-T) function~\cite{Adroja1,Adroja2,Adroja3,Bhattacharyya3},

\begin{equation}
G_{\mathrm{z2}}(t) =A_{\mathrm{0}} G_{\mathrm{KT}}(t)e^{-\lambda t}+A_{\mathrm{bg}}
\end{equation}

where
\begin{equation}
G_{\mathrm{KT}}(t) =\left[\frac{1}{3}+\frac{2}{3}(1-\sigma_{\mathrm{KT}}^2t^2)e^{{\frac{-\sigma_{\mathrm{KT}}^2t^2}{2}}}\right]
\end{equation}

\noindent is the Kubo-Toyabe (KT) functional form expected from an isotropic Gaussian distribution of randomly oriented static (or quasi-static) local fields at muon sites. $\lambda$ is the electronic relaxation rate, $A_{\mathrm{0}}$ is the initial asymmetry, $A_{\mathrm{bg}}$ is the background.  The parameters $\sigma_{\mathrm{KT}}$, $A_{\mathrm{0}}$, and $A_{\mathrm{bg}}$ are found to be temperature independent.  Our zero-field$-\mu$SR measurements indicates that the time-reversal symmetry is preserved in the superconducting state of LaIr$_3$. In Mg$_{10}$Ir$_{19}$B$_{16}$, ZF$-\mu$SR measurements also find no evidence for time-reversal symmetry-breaking fields~\cite{A. A. Aczel}. In the case of La$_7$Ir$_3$, ZF$-\mu$SR measurements reveal the presence of spontaneous static or quasistatic magnetic fields below the superconducting transition temperature suggest that in this case the superconducting state breaks time-reversal symmetry~\cite{J.A.T. Barker}.

\section{Summary}

\noindent In summary, we have examined the characteristics of the superconducting ground state in LaIr$_3$ using TF and ZF$-$muon spin relaxation/rotation measurements. Bulk type-II superconductivity is observed in the susceptibility measurements with $T_{\mathrm{C}}$ = 2.5 K. Similar to UPt$_3$ two consecutive transitions are seen in the heat capacity data and the origin of the lower temperature feature remains unclear at this time. Transverse field$-\mu$SR measurements reveal a fully gapped $s-$wave type superconductivity with the dimensionless ratio, 2$\Delta(0)/k_{\mathrm{B}}T_{\mathrm{C}}$ = 3.31(1), compared to 3.56 (BCS value)  suggesting weak coupling superconductivity. Our zero-field$-\mu$SR measurements do not reveal the spontaneous appearance of an internal magnetic field below the transition temperature, which indicates that the time-reversal symmetry is preserved in the superconducting state. The results underline the need for further research into the properties of Ir-based intermetallic superconductors and especially those that have a noncentrosymmetric structure. Strong spin-orbit coupling in these systems would then be antisymmetric, which often leads to fascinating and unconventional superconductivity with mixed pairing.

\section{Acknowledgments}

AB would like to acknowledge DST India, for Inspire Faculty Research Grant (DST/INSPIRE/04/2015/000169). DTA and ADH would like to thank CMPC-STFC, grant number CMPC-09108, and also DIST for financial support. DTA thanks to JSPS for invitation fellowship.

\end{document}